\documentstyle[prb,epsf,eqsecnum,aps,multicol]{revtex}

\def\beq{\begin{equation}}
\def\eeq{\end{equation}}

\def\prb{Phys. Rev. B }
\def\pra{Phys. Rev. A }
\def\prl{Phys. Rev. Lett. }
\def\ajp{Am. J. Phys. }
\def\mpl{Mod. Phys. Lett. B }
\def\ijmp{Int. J. Mod. Phys. B }
\def\ijp{Ind. J. Phys. }
\def\ibmjrd{IBM J. Res. Dev. }
\def\pjp{Pramana J. Phys. }

\begin{document}

\draft

\title{Dephasing of Aharonov-Bohm oscillations in a mesoscopic ring
  with a magnetic impurity} 

\author{Sandeep K. Joshi\cite{joshie}, Debendranath
    Sahoo\cite{sahooe} and A. M. Jayannavar\cite{amje} } 
 
\address{Institute of Physics, Sachivalaya Marg, Bhubaneswar 751 005,
  Orissa, India} 

\date{\today}

\maketitle

\begin{abstract}
We present a detailed analysis of the Aharonov-Bohm interference
oscillations manifested through transmission of an electron in a
mesoscopic ring with a magnetic impurity atom inserted in one of its arms.
The electron interacts with the impurity through the exchange interaction
leading to exchange spin-flip scattering. Transmission in the spin-flipped
and spin-unflipped channels are explicitly calculated. We show that the
spin-flipper acts as a dephasor in spite of absence of any inelastic
scattering. The spin-conductance (related to spin-polarized transmission
coefficient) is asymmetric in the flux reversal as opposed to the two probe
conductance which is symmetric under flux reversal.
\end{abstract}


\begin{multicols}{2}

\begin{figure}
\protect\centerline{\epsfxsize=2.5in \epsfbox{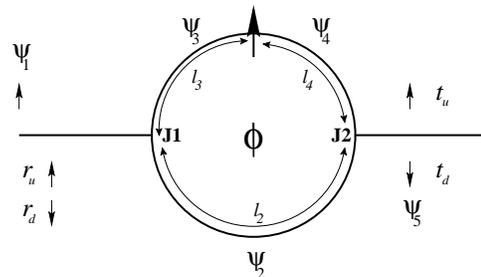}}
\caption{Mesoscopic ring with Aharonov-Bohm flux $\phi$ threading
  through the center of the ring and a magnetic impurity in one arm of
  the ring.}
\label{ring}
\end{figure}

  
  Quantum transport in open mesoscopic systems has attracted
  considerable attention in the last two decades \cite{imry_book,datta,psd}. In
  this area, the study of phase coherent transmission of electrons in
  the Aharonov-Bohm (AB) ring occupies a prominent place
  \cite{imry_book,datta,psd,webb_ap,gia,wgtr}. Study of dephasing \cite{sai,ims_pb,butt_prb,pareek,nku} of electrons in this geometry
  is very timely \cite{shtrikman} to understand basic issues related
  to quantum phenomena. By introducing a magnetic impurity atom (to be
  referred to as the spin-flipper, or the flipper, for short) in one
  arm of the ring, one can couple the spin of the electron
  ($\vec{\sigma}$) to the spin of the flipper ($\vec{S}$) via the
  exchange interaction \cite{imry_book,sai}.  This leads to scattering
  of the electron in which the spin state of the electron and the impurity 
  is changed without any exchange of energy
  leading to reduction of interference. Let the electron be incident
  from the left reservoir with its spin pointing ``up'' (see Fig.
  \ref{ring}). The spin of the electron passing through the upper arm
  may or may not be flipped by the flipper. In the case that the spin
  is unflipped, one would expect the usual AB-oscillations of the
  transmission due to interference of the partial waves passing
  through the upper and the lower branches of the ring. However, in
  the case that the spin is flipped, one would think, guided by naive
  intuition, that a path detection has taken place and hence one would
  be led to conclude that the interference pattern for the spin-down
  component would be wiped out. This is true provided we consider only
  two forward propagating partial waves. However, there are infinitely
  many partial waves in this geometry which are to be superposed to
  get the total transmission. These arise due to the multiple
  reflections from the junctions and the impurity site. Consider, for
  example, an incident spin-up particle moving in the upper arm which
  is flipped at the impurity site and gets reflected to finally
  traverse the lower arm before being transmitted. Naturally, this
  partial wave will interfere with the spin-flipped component
  transmitted along the upper arm. This results in non-zero
  transmission for the spin-flipped electron.  Thus on taking into
  account the multiple reflections (more than just two partial waves)
  the presence of magnetic impurity does not lead to "which-path"
  information.  However, the magnetic impurity acts as a dephasor
  \cite{imry_book,sai}. 

  In this work we show that the amplitude of AB-oscillations is
  reduced by the flipper causing dephasing. We study the problem using
  the quantum waveguide theory approach \cite{wgtr,xia} and the spin
  degree of freedom of the electron is dealt with in line with ref.
  \onlinecite{ajp}. We consider an impurity consisting of a flipper
  capable of existing in M different discrete internal spin states and
  located at a particular position on the upper arm of the ring (see
  Fig. \ref{ring}). The spin $\vec{\sigma}$ of the electron couples to
  the flipper spin $\vec{S}$ via an exchange interaction $-J
  \vec{\sigma} \cdot \vec{S} \delta(x-l_3)$. The magnetic flux
  threading the ring is denoted by $\phi$ and is related to the vector
  potential $A=\phi/l$, $l$ being the ring circumference \cite{xia}.
  During passage of the electron through the ring, the total spin
  angular momentum and its $z$-component remain conserved.  We analyze
  the nature of spin-up/down and total transmission (reflection)
  coefficients. For this we consider the incident electron to be
  spin-polarized in the up-direction.  Apart from dephasing we show
  that up/down-transmission coefficients are asymmetric in flux
  reversal, i.e., total spin polarization (related to spin conductance
  \cite{sarma}) is asymmetric in flux reversal. As expected we find
  that the total transmission coefficient which is the sum of spin-up
  and spin-down transmission coefficients is symmetric in the flux
  reversal.
  
  Let $l_2$ be the length of the lower arm of the ring and the
  impurity atom be placed at a distance $l_3$ from the junction J1,
  $l_4$ being the remaining segment length of the upper arm. The
  various segments of the ring and its leads are labelled as shown in
  Fig. \ref{ring} and the wave functions in these segments carry the
  corresponding subscripts. The wave functions in the five segments
  for a left-incident spin-up electron can be written as follows:

\begin{eqnarray}
\label{eq:1}
\psi _1&=&(e^{ikx}+r_u e^{-ikx})\chi _m\alpha+\nonumber\\
          &&r_d e^{-ikx}\chi _{m+1}\beta,\nonumber\\
\psi _2&=&(A_u e^{ik_1x}+B_u e^{-ik_2x})\chi _m\alpha+\nonumber\\
          &&(A_d e^{ik_1x}+B_d e^{-ik_2x})\chi _{m+1}\beta,\nonumber\\
\psi _3&=&(C_u e^{ik_1x}+D_u e^{-ik_2x})\chi _m\alpha+\nonumber\\
          &&(C_d e^{ik_1x}+D_d e^{-ik_2x})\chi _{m+1}\beta,\nonumber\\
\psi _4&=&(E_u e^{ik_1x}+F_u e^{-ik_2x})\chi _m\alpha+\nonumber\\
          &&(E_d e^{ik_1x}+F_d e^{-ik_2x})\chi _{m+1}\beta,\nonumber\\
\psi _5&=&t_u e^{ikx}\chi _m\alpha+t_d e^{ikx}\chi _{m+1}\beta.
\end{eqnarray}

\noindent where $k_1=k+(e\phi /\hbar cl)$, $k_2=k-(e\phi /\hbar cl)$,
$k$ is the wave-vector of incident electron. The subscripts $u$ and
$d$ represent ``up'' and ``down'' spin states of the electron with the
corresponding spinors $\alpha$ and $\beta$ respectively (i.e., $\sigma
_z\alpha =\frac{1}{2}\alpha$, $\sigma _z\beta =-\frac{1}{2}\beta$) and
$\chi _m$ denotes the wave function of the impurity \cite{ajp} with
$S_z=m$ (i.e., $S_z\chi _m=m\chi _m$). The reflected (transmitted)
waves have amplitudes $r_u$ ($t_u$) and $r_d$ ($t_d$) corresponding to
the ``up'' and ``down'' spin components respectively.  Continuity of
the wave functions and the current conservation\cite{wgtr,xia,ajp} at
the junctions J1 and J2 imply the following boundary conditions.

\begin{eqnarray}
  \label{eq:4}
  \psi _1(x=0)=\psi _2(x=0),\nonumber\\
  \psi _1(x=0)=\psi _3(x=0),\nonumber\\
  \psi _1^\prime (x=0)=\psi _2^\prime (x=0)+\psi _3^\prime (x=0),\nonumber\\
  \psi _3^\prime (x=l_3)-\psi _4^\prime
          (x=l_3)=G(\vec{\sigma}\cdot\vec{S})\psi _3(x=l_3),\nonumber\\
  \psi _3(x=l_3)=\psi_4(x=l_4),\nonumber\\
  \psi _4(x=l_3+l_4)=\psi _5(x=0),\nonumber\\
  \psi _2(x=l_2)=\psi _5(x=0),\nonumber\\
  \psi _2^\prime (x=l_2)+\psi _4^\prime (x=l_3+l_4)=\psi _5^\prime (x=0).
\end{eqnarray}

\noindent Here $G=2mJ/\hbar ^2$ is the coupling constant indicative of the
``strength'' of the spin-exchange interaction.  The primes denote the
spatial derivatives of the wave functions. Equations (\ref{eq:1}) along
with the boundary conditions (\ref{eq:4}) were solved to obtain the
amplitudes $t_u$, $t_d$, $r_u$ and $r_d$. Owing to the large length of the
expressions in the following we confine ourselves to the graphical
interpretation of the results. We have taken the flipper to be a spin-half
object ($M=2$) situated symmetrically at the center of the upper arm, i.e.,
$l_3=l_4$. Now, depending upon the initial state of the flipper we have
possibility of either spin-flip scattering ($\sigma_z=1/2,~S_z=-1/2$) or
no spin-flip scattering ($\sigma_z=1/2,~S_z=1/2$), as demanded by the
conservation of the total spin and its $z$-component. In the case of
no-spin-flip scattering ($\sigma_z=1/2,~S_z=1/2$) the problem at hand
reduces to that of simple potential scattering from the impurity. We have 
set $\hbar=2m=1$ and throughout the value of interaction strength $G$ is 
given in dimensionless units. The parameters used for the analysis are
mentioned in the figure captions.


\begin{figure}
\protect\centerline{\epsfxsize=2.5in \epsfbox{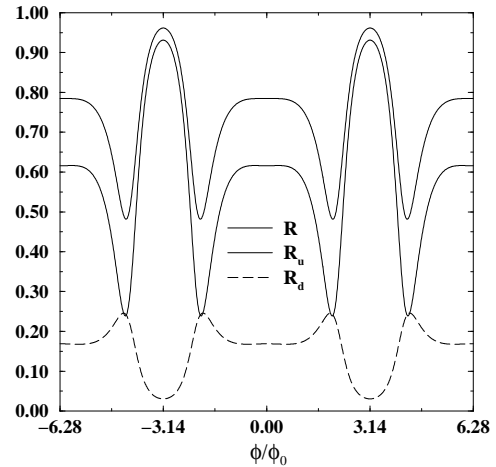}}
\caption{ Plot of total reflection coefficient $R$, spin-up reflection
coefficient $R_u$ and spin-down reflection coefficient $R_d$ for the spin
-flip scattering case. The parameters are $kl=1.0$, $G=10.0$.  }
\label{symm-refl}
\end{figure}

\begin{figure}
\protect\centerline{\epsfxsize=2.5in \epsfbox{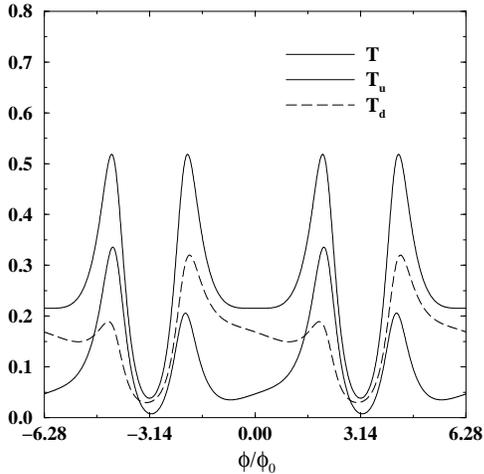}}
\caption{ Plot of total transmission coefficient $T$, spin-up transmission
coefficient $T_u$ and spin-down transmission coefficient $T_d$ for the spin
-flip scattering case. The parameters are $kl=1.0$, $G=10.0$.  }
\label{symm-trans}
\end{figure}

To begin with we take a look at the symmetry properties of the
transport coefficients in spin-flip scattering case where the electron
spin is opposite to the flipper spin. It is worth noting that due to
the presence of spin degree of freedom the problem in hand although
one-dimensional becomes a multichannel problem. Figure \ref{symm-refl}
shows the spin-up reflection coefficient $R_u=|r_u|^2$, spin-down
reflection coefficient $R_d=|r_d|^2$ and total reflection coefficient
$R=R_u+R_d$ as a function of the magnetic flux parameter
$\eta=\phi/\phi_0$, $\phi_0$ being the flux quantum $hc/e$. We clearly
see the AB-oscillations with flux periodicity\cite{webb_ap} of $2\pi\phi_0$. All
three reflection coefficients are symmetric in the flux reversal as
expected on general grounds\cite{butt_ibm}. In Fig.
\ref{symm-trans} we plot the spin-up transmission coefficient
$T_u=|t_u|^2$ (thin line), spin-down transmission coefficient
$T_d=|t_d|^2$ (dashed line) and total transmission coefficient
$T=T_u+T_d$ (thick line) versus $\eta$. It unambiguously shows
that though the total transmission $T$ (related to the two-terminal
electrical conductance) is symmetric in flux reversal the spin-up
$T_u$ and spin-down $T_d$ components are asymmetric under flux
reversal. These transmission coefficients show AB-oscillations with
flux periodicity of $2\pi\phi_0$.  We have verified this behavior of
the reflection and transmission coefficients for various values of
wave vector $k$ of the incident electron and impurity strength $G$.
These observations are consistent with the reciprocity relations for
transport in multichannel systems \cite{butt_ibm} and are
a consequence of the general symmetry properties of the Hamiltonian \cite{datta}.
The transmission coefficient at flux $\phi$ for the case when the
incident particle is spin-up and the impurity is spin-down is equal to
the transmission coefficient for the case when incident particle is
spin-down and impurity is spin-up but the flux direction is reversed.
For the spin-polarized transport the total polarization $T_u-T_d$ is
related to the spin-conductance \cite{sarma}. The above symmetry
properties imply that the spin-conductance is asymmetric under the
flux reversal. It should be noted that at zero temperature the total
electrical conductance is calculated by summing up with equal
weight-age the total transmission coefficients for all the four cases,
i.e., $\sigma_z=\pm 1/2$ and $S_z=\pm 1/2$.

\begin{figure}
\protect\centerline{\epsfxsize=2.5in \epsfbox{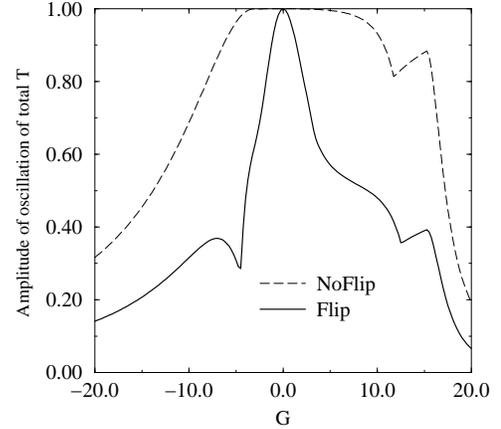}}
\caption{ Variation of amplitude of oscillation of total transmission
coefficient with the interaction strength for the two cases of flip
and no-flip scattering. The parameters are $kl=1.0$ and $\eta=1.0 $. }
\label{ampli-reduc}
\end{figure}

As discussed in the introduction, due to multiple reflections the presence
of a spin-flipper in one arm does not lead to "which-path" information.
This would have implied the complete blocking of spin-down transmission.
In contrast we clearly observe the AB-oscillations for the case of $T_d$
originating from multiple reflections. We now address the question of
dephasing due to the spin-flipper. To quantify dephasing , we calculate
the amplitude of AB oscillations (also referred to as visibility factor)
by taking the difference between the maximum and the minimum of total
transmission coefficient as a function of flux $\phi$ over one period of
the oscillation. A plot of the variation of the amplitude of oscillation
of total transmission $T$ with the interaction strength $G$ for the two
cases, no spin-flip scattering ($S=1/2~m=1/2$) and spin-flip scattering
($S=1/2~m=-1/2$), is shown in figure Fig.\ref{ampli-reduc} with dashed line
and thick line respectively. Note, however, the signature of dephasing is
that the amplitude of AB oscillation of transmission coefficient for the
spin-flip case is always smaller than that for the no spin-flip case for
all non-zero values of coupling strength $G$. In other words the reduction
of amplitude of AB oscillations is stronger for the spin-flip scattering
case. We have verified the above observation for other parameters in the
problem. Thus the presence of spin-flipper reduces or dephases the
AB-oscillations and it acts as a dephasor\cite{imry_book,sai,ims_pb}. 

\begin{figure}
\protect\centerline{\epsfxsize=2.5in \epsfbox{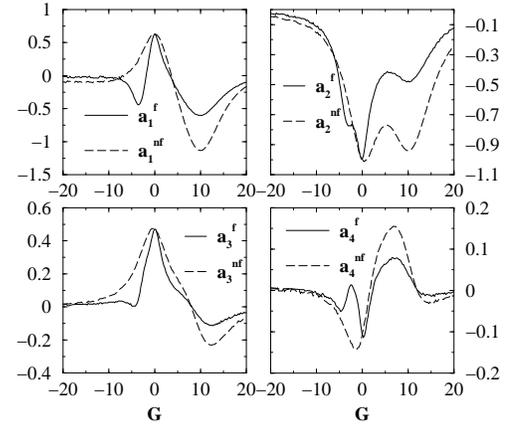}}
\caption{ Variation of $n^{th}$ harmonic component $a_n$ of the total
transmission coefficient with the coupling strength $G$ at $kl=1.0$. 
Dashed lines are for the no-flip case and solid lines are for the
flip case.}
\label{fourier}
\end{figure}
 
At this point we are inclined to think that the harmonic components of
the total transmission $T(\eta)$ in $\eta=\phi/\phi_0$ might be
able to shed more light on the issue. So, with the hope of extracting
some systematics we plot the $n^{th}$ harmonic component 
$a_n=\int_0^{2\pi} T(\eta) cos(n \eta) d \eta$ for $n=1,2,3...$ as a
function of strength $G$ for the spin-flip scattering as well as no
spin-flip scattering cases. The plots are shown in Fig. \ref{fourier}
for first four harmonic components.
As can be seen the harmonic components do not show any systematics in
the sense that the higher harmonic components can dominate over the
lower harmonic components at certain values of strength $G$ for
spin-flip scattering as well as no spin-flip scattering cases. {\em
  Also, the} $n^{th}$ {\em harmonic component for spin-flip scattering
  could dominate over that of the no spin-flip scattering component}.
These features of the harmonic components are manifestations of the
multiple scattering nature of the transport in such ballistic systems
as against the observation of domination of lowest harmonic component
($n=1$) in the case of transport in the presence of evanescent modes
\cite{ev_modes}. Guided by the naive intuition mentioned earlier we
would have expected the lowest harmonic to dominate. This reiterates
the important role played by the reflection at the impurity site. We
would like to emphasize that irrespective of the behavior of the
harmonic components (say for a particular case $n^{th}$ harmonic
component in the spin-flip case is dominant over the same $n^{th}$
harmonic component for no-spin-flip case) the AB-oscillations of the
total transmission are always dephased in the spin-flip case.

\begin{figure}
\protect\centerline{\epsfxsize=2.5in \epsfbox{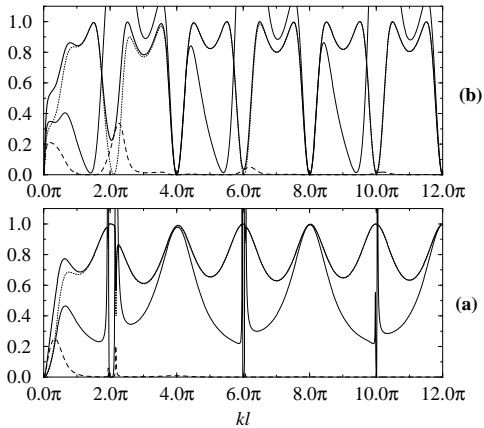}}
\caption{ The transmission spectrum of $T$(thick line), $T_u$(dotted line)
          and $T_d$(dashed line) for
          (a) $\eta=0.0$ $G=5.0$ and (b) $\eta=1.3$,$G=5.0$. Thin
          line shows the plot of the unnormalized electron probability
          $|\psi_3(l_3)|^2$ at the impurity site $x=l3$.}
\label{kspec}
\end{figure}

In order to make sure that nothing unusual happens at other energies we
study the $T$, $T_u$ and $T_d$ as functions of
$kl$ for the case of spin-flip scattering.
Figure \ref{kspec} reveals an interesting fact, namely at
$kl=2\pi+4n\pi,~n=0,1,2...$ the $T_d$ component vanishes independent of
the value of interaction strength $G$. In the $\eta\neq 0$ case this
happens at $kl=4n\pi,~n=1,2,...$. At these values of the incident
wave-vectors the electron wave function at the impurity site happens to be
zero. As a result the electron does not interact with the impurity at all
and consequently there is no spin-flip scattering at these energies.
However, these k-points are to be distinguished from those at which
although $T_d$ is zero but in addition $T_u=1$, because at these resonant
energies the restriction $T+R=1$ forces $T_d,R_u$ and $R_d$ to be zero.

In conclusion, we have studied in detail the nature of AB-oscillations in
mesoscopic ring in the presence of a spin-flipper in one of its arms. We
have shown that this acts as a dephasor. The presence of magnetic impurity
makes the polarized transmission coefficient asymmetric in flux reversal
whereas the total transmission coefficient is symmetric in line with the
theoretical expectations. Further case of asymmetrically placed flipper,
spin-flipper with higher number of internal states and spin-flippers in
both arms of the ring are under investigation for studying the nature of
resonances, phase-shifts and dephasing.


\end{multicols}

\end{document}